\documentclass[preprint,sort&compress,5p]{elsarticle}

\usepackage{hyperref,amsmath}
\usepackage{color}

\journal{Advances in Colloid and Interface Science}

\begin{document}

\begin{frontmatter}

\title{Magnetocapillary self-assemblies: locomotion and micromanipulation along a liquid interface}

\author{G. Grosjean}
\ead{ga.grosjean@ulg.ac.be}
\author{M. Hubert, N. Vandewalle}
\address{Universit\'e de Li\`ege, GRASP Lab, CESAM Research Unit, All\'ee du 6 Ao\^ut 19, 4000 Li\`ege, Belgium}

\begin{abstract}
This paper presents an overview and discussion of magnetocapillary self-assemblies.
New results are presented, in particular concerning the possible development of future applications.
These self-organizing structures possess the notable ability to move along an interface when powered by an oscillatory, uniform magnetic field.
The system is constructed as follows.
Soft magnetic particles are placed on a liquid interface, and submitted to a magnetic induction field.
An attractive force due to the curvature of the interface around the particles competes with an interaction between magnetic dipoles.
Ordered structures can spontaneously emerge from these conditions.
Furthermore, time-dependent magnetic fields can produce a wide range of dynamic behaviours, including non-time-reversible deformation sequences that produce translational motion at low Reynolds number.
In other words, due to a spontaneous breaking of time-reversal symmetry, the assembly can turn into a surface microswimmer.
Trajectories have been shown to be precisely controllable.
As a consequence, this system offers a way to produce microrobots able to perform different tasks.
This is illustrated in this paper by the capture, transport and release of a floating cargo, and the controlled mixing of fluids at low Reynolds number.
\end{abstract}

\begin{keyword}
Microswimmers \sep Self-assembly \sep Capillary forces
\end{keyword}

\end{frontmatter}


\section{Introduction}

Particles floating at an interface can interact and form aggregates~\cite{nicolson1949,kralchevsky1994,kralchevsky2000,vella2005}.
This is a well-known effect of the deformation of the surface around each particle.
Indeed, a single floating particle is often surrounded by a meniscus that is a function of its shape, buoyancy, and wetting properties~\cite{kralchevsky1994,danov2010,poty2014}.
For instance, a heavy and/or hydrophobic sphere will create a concave depression on a water surface.
When floating on a sloped surface, it can experience a lateral force, as it minimizes its potential energy~\cite{nicolson1949}.
Depending on whether the particle is buoyant or not, it will tend to move up or down the slope, respectively.
Two neighbouring particles will therefore attract or repel, as each particle experiences an inclination of the interface caused by the other~\cite{kralchevsky1994}.
Two heavy spheres will thus attract and cluster, as they are each dragged by gravity in the depression around the other.
This phenomenon of agglomeration is sometimes colloquially known as the Cheerios effect, as it can be observed with breakfast cereals floating in a bowl of milk~\cite{vella2005}.

It has been proposed to use magnetic floating particles to generate self-assemblies~\cite{golosovsky1999,wen2000,grzybowski2000,golosovsky2002,snezhko2006,vandewalle2012,vandewalle2013}.
When a magnetisation is induced perpendicular to the interface, the particles experience a repulsive dipole-dipole interaction at short range, opposing the capillary attraction and preventing clustering~\cite{wen2000,vandewalle2012,vandewalle2013}.
This can lead to the appearance of a finite equilibrium distance, so that the particles self-organize into a triangular lattice, which would be the expected symmetry on a flat surface~\cite{messina2015}.
Other planar crystal symmetries have also been observed~\cite{wen2000} as well as defects, such as fivefold symmetries in polydisperse assemblies~\cite{wen2000} or in heavy assemblies, when a large curvature of the surface is reached due to the combined weight of the particles~\cite{vandewalle2012,vandewalle2013}.
Similar structures have been reached with little to no capillary interaction, using a non-uniform magnetic field for confinement instead~\cite{golosovsky1999,golosovsky2002}.
Floating magnetic particles can also form structures outside of thermodynamic equilibrium, called dynamic self-assemblies~\cite{grzybowski2000,snezhko2006,snezhko2011}.
Such structures rely on a constant supply of energy, for instance using time-dependent magnetic fields.
Examples of forces driving these dynamic assemblies include self-induced surface waves~\cite{snezhko2006,snezhko2011}, and hydrodynamic interactions induced by the rapid rotation of the particles~\cite{grzybowski2000}.

This paper focuses on recent developments in the study of few-body assemblies of metallic spheres on a water surface.
Typically, spheres of diameter $D \approx 500~\mathrm{\mu m}$ are used, so that thermal agitation is negligible.
Contact between particles is usually prevented, as they are exposed to a constant vertical magnetic induction field $B_z$ of the order of a few mT.
A time-dependent horizontal magnetic field $\vec{B}_h (t) = B_x (t) \vec{e}_x + B_y (t) \vec{e}_y$ is added to generate deformations in the self-assembly.

\section{Methods}

Metallic spheres are placed on a water bath.
The steels or alloys used (mainly UNS S42000 and G52986) and the spherical shape of the particles allow for a linear magnetisation, with little residual magnetism~\cite{lagubeau2016}.
The magnetic moment of a particle exposed to an induction field $\vec{B}$ is expressed by $\vec{m} = \chi V \vec{B}/\mu_0$, where $\mu_0$ is the vacuum permeability, $V$ the volume of a sphere and $\chi$ its effective susceptibility.
For a spherical object, it is linked to the bulk susceptibility $\chi_{\mathrm{bulk}}$ by the relation $\chi = \chi_{\mathrm{bulk}}/(1+\chi_{\mathrm{bulk}}/3)$, such that for materials with a large susceptibility, we have $\chi \approx 3$~\cite{osborn1945}.
The materials used herein have a bulk susceptibility $\chi_{\mathrm{bulk}} > 300$.
The magnetic dipole-dipole potential between two identical spheres separated by a distance $d$ is given by
\begin{equation}
U_m = \frac{\mu_0 \left[ m_z^2 + m_h^2 \left(1 - 3\cos^2 \theta \right)\right]}{4\pi d^3},
\label{Um}
\end{equation}
where $\theta$ is the angle between the relative position of the pair and the horizontal magnetic field $\vec{B}_h$.
Vertical and horizontal components of the magnetic moments $m_z$ and $m_h$ have been separated, as $m_z$ can only generate a repulsion and is usually kept constant in the course of an experiment.

\begin{figure}
\includegraphics[width=\linewidth]{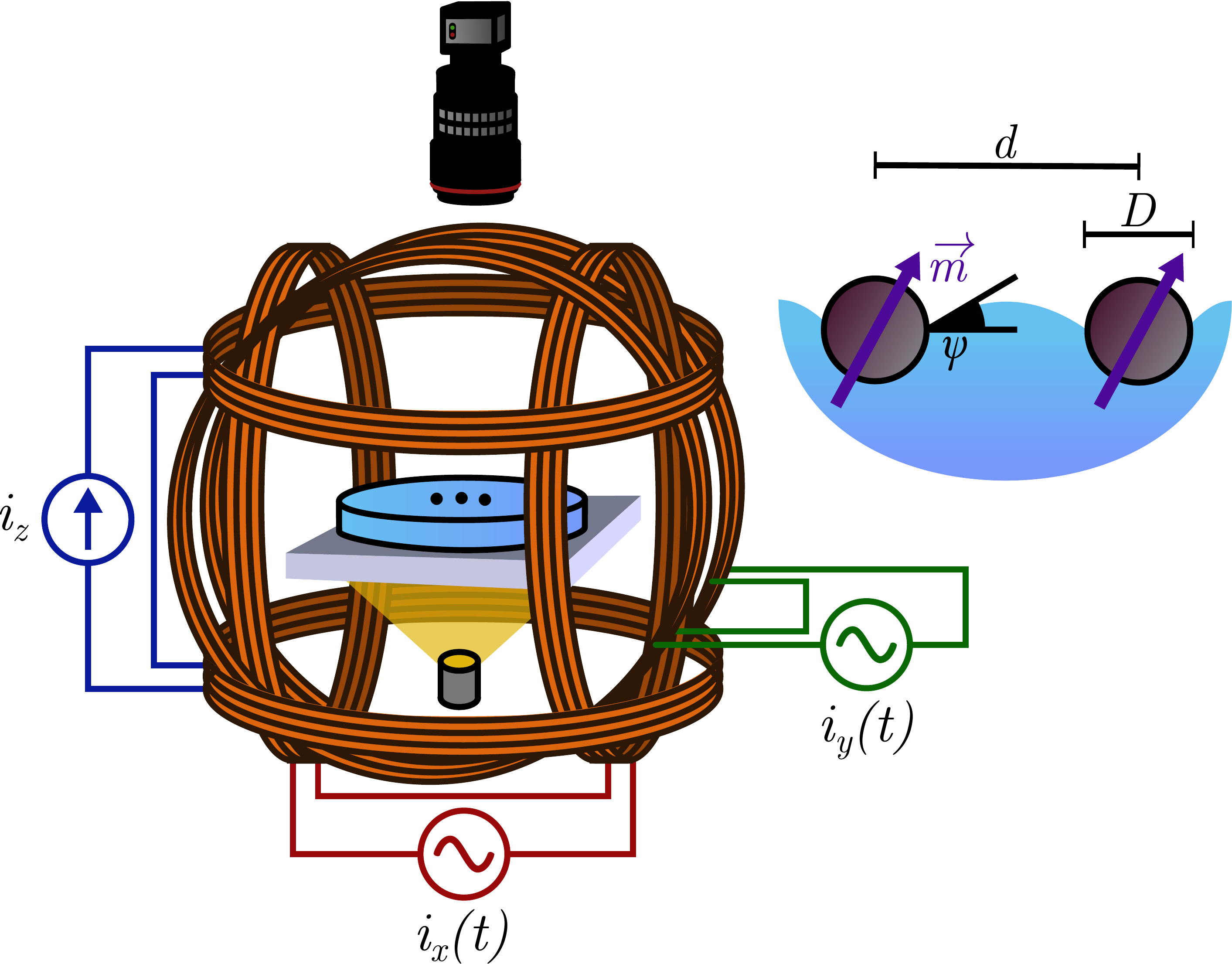}
\caption{
Experimental setup of magnetocapillary self-assemblies.
A water bath is placed in the center of a tri-axis Helmholtz coil system.
A direct current $i_z$ is injected in the $z$ coils to generate a constant vertical magnetic induction field $B_z$.
Currents $i_x (t)$ and $i_y (t)$ are injected in the horizontal coils to generate time-dependent horizontal magnetic fields $B_x (t)$ and $B_y (t)$.
Steel spheres of diameter $D$ floating on the bath possess a magnetic induced moment $\vec{m}$ proportional to the total magnetic field in the center of the coils.}
\label{setup}
\end{figure}

Figure~\ref{setup} illustrates the experimental setup.
The magnetic induction fields are generated using three orthogonal pairs of Helmholtz coils.
A direct current power supply provides the $z$ coils with a current $i_z$.
A multichannel arbitrary function generator going through a pairs of linear amplifiers feeds the $x$ and $y$ coils with currents $i_x (t)$ and $i_y (t)$.
The water bath is at the center of the coils.
It is lit from below through a diffuser and filmed from atop using a CCD video camera and a macro lens.

The capillary interaction potential can be simplified by the assumption that the deformation of the surface in the presence of two particles is the sum of the deformations caused by each particle individually.
This approximation holds for a low number of particles separated by a large distance $d \gg D$~\cite{lagubeau2016}.
For small deformations, the meniscus profile around a particle is given by the Laplace equation and reads
\begin{equation}
\frac{2z}{D} = A K_0 (d/l_c),
\label{zeta}
\end{equation}
where $A$ is given by the boundary conditions at the contact line, $l_c$ is the capillary length and $K_0$ is the zero order modified Bessel function of the second kind~\cite{osborn1945}.
The potential energy associated with the capillary interaction between two identical spheres is therefore given by
\begin{equation}
U_c = -2\pi\gamma q^2 K_0 (d/l_c),
\label{Uc}
\end{equation}
where $\gamma$ denotes the surface tension and $q = a \sin \psi$ is a typical deformation length called the capillary charge, by analogy with the Coulomb interaction potential~\cite{kralchevsky1994}.
It is a function of the contact line radius $a$ and the meniscus slope angle $\psi$.
Contrary to electric charges, like capillary charges attract and unlike charges repel.

Both materials used have a density $\rho_s$ of approximately 7800~$\mathrm{kg/m^3}$ and thus would not float without surface tension.
This places an upper bound on the diameter $D$ of the spheres around $1~\mathrm{mm}$ on water. 
For most experiments, we have monodisperse assemblies with $D=397$, $500$ or $793~\mathrm{\mu m}$, meaning that thermal agitation is considered negligible.
At room temperature, the diameter under which thermal agitation overcomes the capillary attraction for a pair of particles is $3.4~\mathrm{\mu m}$, which gives a lower bound on $D$~\cite{lagubeau2016}.
However, this bound could be further lowered by relying on other forces than gravity to generate the surface deformation, such as wetting~\cite{kralchevsky1994} or a magnetic force~\cite{vella2015}, or by using another geometry for the interface~\cite{kralchevsky2000,ershov2013}.

One can compare the magnetic and capillary energies by defining a magnetocapillary number~\cite{lagubeau2016,chinomona2015}.
Let $\mathcal{M}(B)$ denote the magnetocapillary number associated with a given magnetic induction field of amplitude $B$.
We have
\begin{equation}
\mathcal{M}(B) = \frac{\chi^2 V^2 B^2}{8\pi^2\gamma q^2 l_c^3 \mu_0}.
\label{M}
\end{equation}
Using this expression, we can write the potential energy of interaction for a pair of particles as
\begin{equation}
\begin{split}
U &= U_m + U_c \\
&= \Gamma \left[ \frac{\mathcal{M}(B_z) + \mathcal{M}(B_h) \left( 1 - 3\cos^2 \theta \right)}{d^3/l_c^3} -K_0 \left(d/l_c\right) \right]
\end{split}
\label{U}
\end{equation}
where we defined $\Gamma=2\pi\gamma q^2$. When $B_h = 0$, the potential shows a competition between the attractive capillary force and a repulsive dipole-dipole interaction.
When a horizontal component $B_h$ is added, a preferential orientation appears, as the interaction energy is minimal for $\theta = 0$, \emph{i.e.} when the pair is aligned with $B_h$.
Furthermore, the contribution of $B_h$ leads to either an attractive or a repulsive force depending on the sign of $\left( 1-3\cos^2\theta \right)$.

\section{Static assemblies}

Depending on the values of $B_z$, $B_h$ and the initial conditions, a wide variety of configurations can be observed.
When $B_h = 0$ and granted that $B_z$ is large enough to overcome the capillary attraction, a triangular lattice is typically observed.
If the weight of the assembly is enough to significantly curve the interface, fivefold symmetries can also be observed~\cite{vandewalle2012,vandewalle2013}.
The addition of a horizontal field $B_h$ can significantly change the symmetry of the assemblies.
Different configurations can coexist, and hysteresis is observed~\cite{vandewalle2013}.
In particular, contact between particles is not always reversible, as a capillary bridge can form between two spheres, significantly increasing the energy required to separate them again.

In the case of a pair of particles, the minimum of Eq.~(\ref{U}) leads to an expression for the equilibrium distance $d_{\mathrm{eq}}$.
We have
\begin{equation}
\frac{d^4_{\mathrm{eq}}}{l^4_c} K_1 (d_{\mathrm{eq}} / l_c) = 3 \mathcal{M}(B_z) - 6 \mathcal{M}(B_h)
\label{deq}
\end{equation}
as well as $\theta = 0$ if $B_h \neq 0$.
A simpler expression can be obtained by considering that $K_1 (d/l_c) \approx l_c/d$, which holds for $d\ll l_c$.
This leads to
\begin{equation}
\frac{d_{\mathrm{eq}}^3}{l_c^3} = 3 \mathcal{M}(B_z) - 6 \mathcal{M}(B_h).
\label{deq2}
\end{equation}
When $d_{\mathrm{eq}} = D$, the particles come into contact.
By injecting this into Eq.~(\ref{deq2}), we find a critical ratio of $B_h$ and $B_z$ at which contact occurs, namely
\begin{equation}
\frac{B_h}{B_z} = \sqrt{\frac{\mathcal{M}(B_h)}{\mathcal{M}(B_z)}} = \sqrt{\frac{1}{2} \left( 1-\frac{D^3}{l_c^3 \mathcal{M}(B_z)} \right)}.
\label{collapse}
\end{equation}
For large values of $\mathcal{M}(B_z)$, this corresponds to $B_h \approx B_z / \sqrt{2}$.
In general, the larger the assembly, the earlier contact occurs.
This is due to two distinct effects.

First, the addition of a horizontal field $B_h$ can either decrease or increase the distance between two particles, depending on $\theta$, the orientation of the pair relative to $B_h$.
If the pair is in line with $B_h$, the horizontal component of the dipole-dipole interaction is an attraction.
Therefore, the distance between particles in a line of 3 or more is further lowered by $B_h$.
For instance, 3 particles can assemble into either a triangle or a line.
The collinear configuration is stable for large values of $B_h$ and $\mathcal{M}_z$, as will be shown later.
In this case, as demonstrated in~\cite{chinomona2015}, contact occurs for
\begin{equation}
\frac{B_h}{B_z} = \sqrt{\frac{1}{2} \left( 1-\frac{24}{17}\frac{D^3}{l_c^3 \mathcal{M}(B_z)} \right)},
\label{collapse3}
\end{equation}
which is slightly lower than the critical value for a pair of particles.

Secondly, the increased weight of the assembly creates a curvature of the interface, lowering the distance between neighbouring particles at the center of the assembly.
This means that, in given conditions, there exists a maximum number of particles $N_c$ above which contact will occur.
As shown in~\cite{vandewalle2013}, by comparing the total weight of the assembly with the capillary force on the boundary of the assembly, a scaling of $N_c$ is found, namely
\begin{equation}
N_c \propto \frac{\gamma^2}{(\rho_s-\rho_l)^2 g^2 D^4}
\label{nmax}
\end{equation}
where $\rho_s$ and $\rho_l$ denote the densities of the particles and the liquid, respectively.
This global curvature is also at the origin of fivefold defects in the assemblies, as a triangular lattice would be expected on a flat surface.

\begin{figure}[t]
\includegraphics[width=\linewidth]{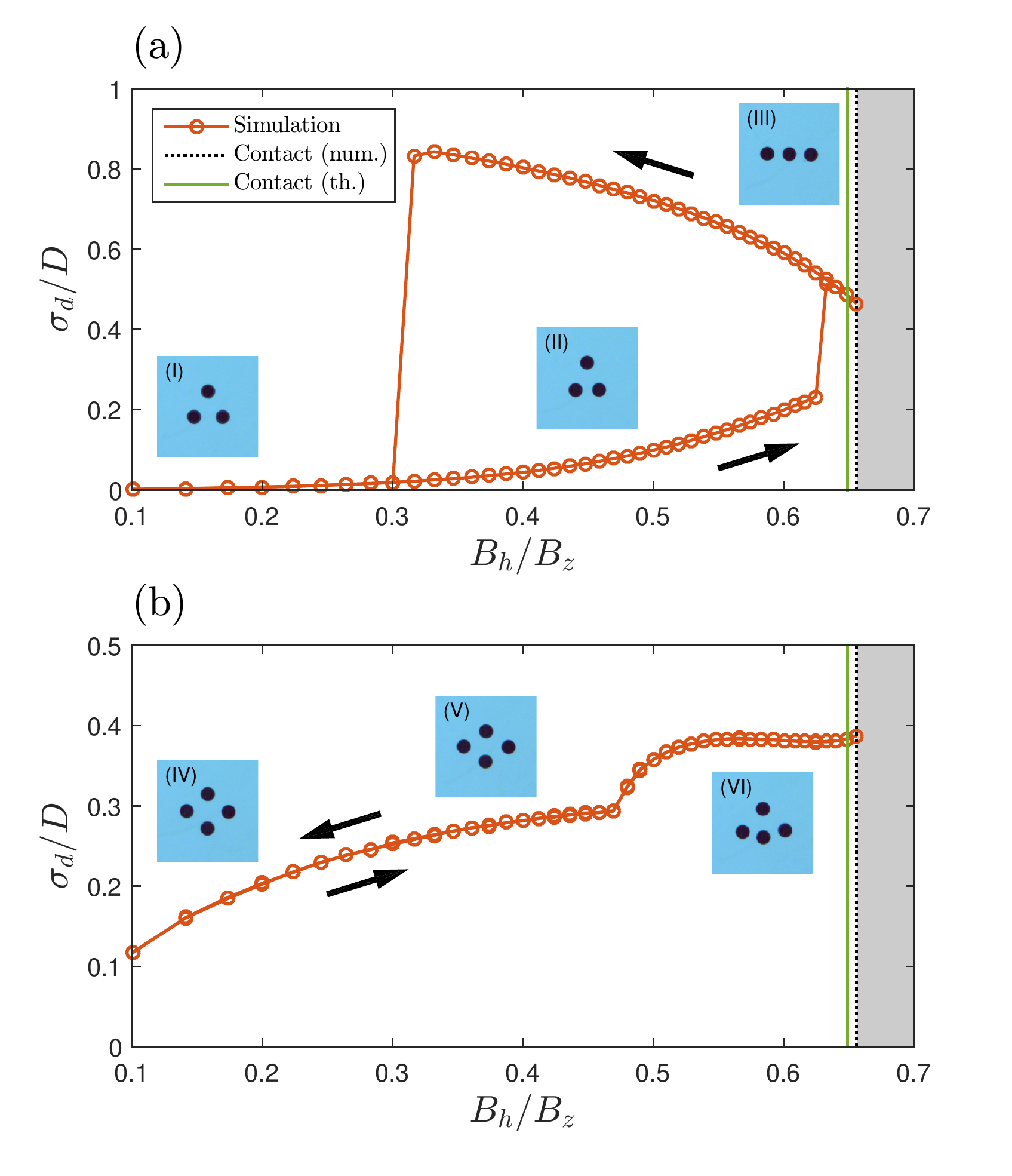}
\caption{
Standard deviation of the distances between the particles and the center of mass during a quasistatic cycle of $B_h/B_z$, from zero to the contact point and back to zero.
The points are obtained by a Monte-Carlo simulation, while the corresponding configurations are shown in the experiment.
The green vertical solid line shows the contact threshold calculated from Eq.~(\ref{collapse3}).
(a) The regular triangle (I) formed by 3 particles deforms into an isosceles (II), then transitions into a collinear state (III).
Hysteresis is observed.
(b) The rhombus (IV) formed by 4 particles continuously deforms (V), then experiences a breaking of mirror symmetry in the direction perpendicular to the field (VI), with no hysteresis.
}
\label{configs}
\end{figure}

As mentioned earlier, a horizontal field can drastically change the symmetry of an assembly.
Figure~\ref{configs} shows the structures obtained as a function of the ratio $B_h/B_z$, for assemblies of 3 and 4 particles.
Different structures are identified by plotting the standard deviation $\sigma_d$ of the distances between the particles and the center of mass, or
\begin{equation}
\frac{\sigma_d}{D} = \sqrt{\frac{1}{N} \sum_{i=1}^{N} \frac{(d_i-\overline{d})^2}{D^2} }
\label{sigma}
\end{equation}
where $d_i$ denotes the distance between particle $i$ and the center of mass and $\overline{d}$ is the average distance to the center of mass.
The curve was obtained through a Monte-Carlo simulation based on the pair potential of Eq.~(\ref{U}).
Horizontal field $B_h$ is increased quasistatically from 0 to the point of contact, then decreased to zero, while $B_z$ is kept constant at about 3~mT.
The insets show the structures obtained in the experiment, next to the corresponding branch in the simulation. 
While contact between particles is generally not reversible in the experiment, the same configurations and hysteretic behaviours are observed.

In the 3-particle case, the equilibrium configuration at $B_h = 0$ is a regular triangle.
When $B_h/B_z$ is increased, the triangle gradually deforms into an isosceles.
Close to the contact event, the system transitions into a collinear configuration.
Contact between particles occurs around $B_h/B_z \approx 0.65$, which is very close to the value obtained from Eq.~(\ref{collapse3}).
When $B_h/B_z$ is decreased, the system remains in the collinear configuration down to $B_h/B_z \approx 0.3$, where it goes back to the isosceles configuration.
This hysteresis loop demonstrates the coexistence of two stable configurations.

On the other hand, no hysteresis is observed in the 4-particle case.
At $B_h/B_z = 0$, we have $\sigma_d  /D \neq 0$.
Indeed, the particles are not equidistant to the center of mass, as the assembly takes the shape of a rhombus with two particles closer to the center, and the other two further away.
When $B_h/B_z$ is increased, this difference becomes larger.
Around $B_h/B_z \approx 0.5$, the symmetry is further broken in the system when the two outermost particles align with one of the innermost ones.
This transition happens continuously.
Contact occurs when the three aligned particles touch, which also happens at $B_h/B_z \approx 0.65$.

Note that representing the deformation of the assemblies with a single variable, $\sigma_d$, does not allow to distinguish configurations that are too similar.
For instance, two isosceles configurations can exist in the 3-particle case, depending on the initial conditions.
However, they both produce similar values of $\sigma_d$.
To distinguish these isosceles requires a more detailed study of the internal angles of the triangular states~\cite{grosjean2015}.
Nonetheless, this brief quasistatic analysis of the assemblies demonstrates the presence of several states with hysteresis loops, and continuous as well as discontinuous transitions.
More importantly, the effect of $N$ on the configurations is profound, meaning that what is known for assemblies of $N$ particles cannot, in general, be transcribed to assemblies of $N'\neq N$ particles.
Each $N$ possesses its own energy landscape, so that the addition or removal of a particle has non-trivial consequences.

\section{Dynamic regimes}

Under time-dependent magnetic fields, magnetocapillary self-assemblies oscillate, producing a wide range of behaviours~\cite{lagubeau2016,chinomona2015,lumay2013,grosjean2015,grosjean2016}.
Most notably, some vibration modes are non-reciprocal, meaning that the succession of shapes adopted by the assembly is not invariant under a time-reversal transformation.
This can lead to the locomotion of the assembly, as was first reported in~\cite{lumay2013}.
The submillimetric size of the particles associated with the low frequencies, usually below 1~Hz, used for the magnetic oscillations lead to a Reynolds number $\mathrm{Re} = \rho_f U D / \eta$ typically between $10^{-3}$ and $10^{-1}$ in water.
This means that, in most cases, viscous dissipations dominate over inertia in the flows produced by the swimmer.
In these conditions, a breaking of time-reversal symmetry in the succession of shapes adopted by the swimmer is a necessary condition for propulsion~\cite{purcell1977,lauga2009}.
Note that this non-reciprocal motion spontaneously arises in the self-assemblies, even under reciprocal perturbations, due to the complex interactions between the particles.

Such a symmetry breaking can already be observed in a pair of particles, although no net motion is observed~\cite{lagubeau2016}.
An oscillating field $\vec{B}_h (t) = B_{x,\:0} \vec{e}_x + B \sin(\omega t) \vec{e}$ is applied to a pair of particles, with $B \ll B_{x,\:0}$ and $\vec{e}$ a unit vector at an angle $\alpha$ with $\vec{e}_x$.
Two vibration modes have been identified, a radial mode in $d$ and an angular mode in $\theta$, corresponding to the cases where the oscillation of the magnetic field is in line with and perpendicular to the pair, \emph{i.e.} $\alpha = 0$ and $\alpha = \pi /2$.
A perturbation analysis of Eq.~(\ref{U}) leads to an expression for the linear stiffness of $U$ in each case, namely
\begin{equation}
k_d = \frac{2 \Gamma}{l_c d}\left( 3 K_1 (d/l_c) - \frac{d}{l_c} K_0 (d/l_c) \right),
\label{kd}
\end{equation}
\begin{equation}
k_{\theta} = \frac{4 \Gamma l_c^3}{d^5}\mathcal{M}(B_x).
\label{ktheta}
\end{equation}
Both modes can therefore be treated as two independent damped oscillators which can experience a forcing by the external magnetic field.
Furthermore, the difference in stiffness means that each oscillation can respond with a different phase to an external oscillating perturbation, producing non-reciprocal deformations. 
However, in the simple two-particle case, the particles oscillate symmetrically around the center of mass of the assembly, leading to no propulsion~\cite{lagubeau2016,lumay2013}.

\begin{figure}
\includegraphics[width=\linewidth]{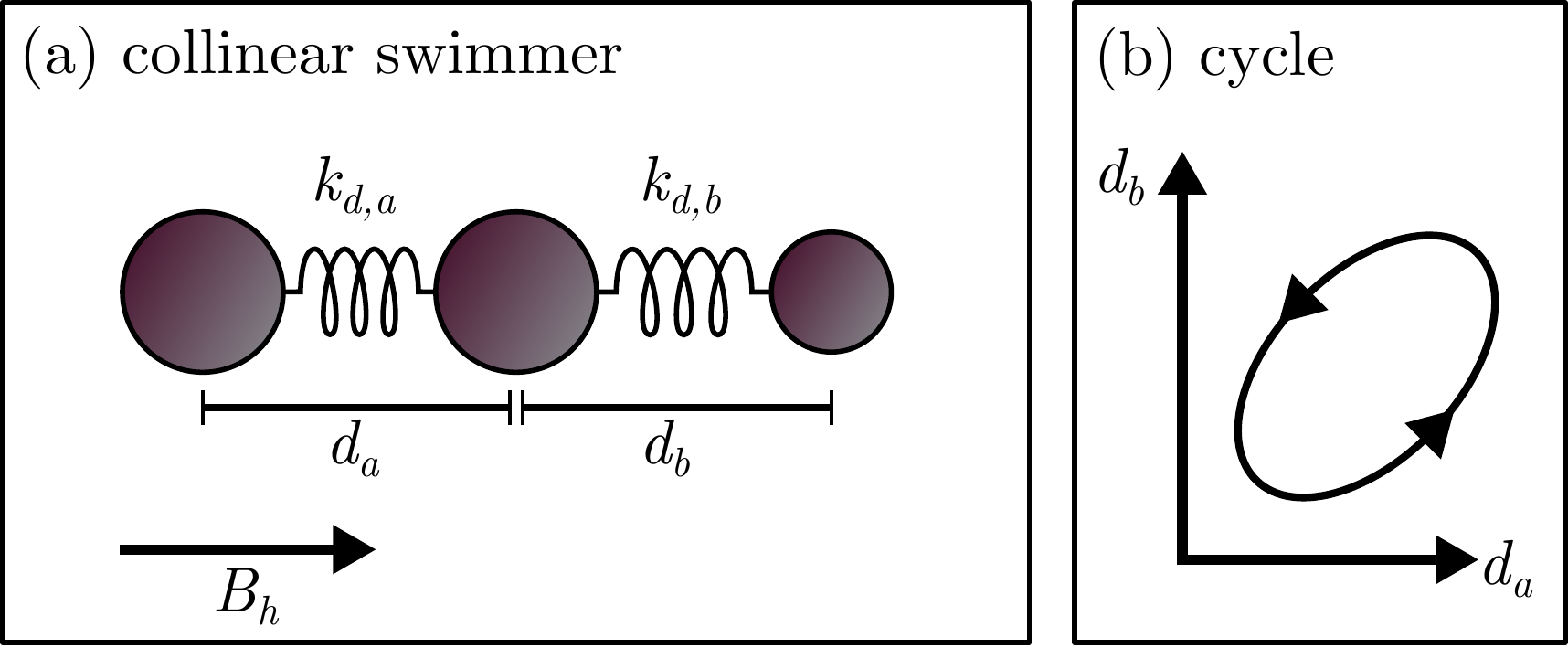}
\caption{
The collinear swimmer is the simplest magnetocapillary swimmer.
(a) Three particles assemble in a line.
One can associate a spring constant with the radial oscillation of each pair, namely $k_a$ and $k_b$.
If the outermost particles differ in size, we have $k_a \neq k_b$.
(b) This can lead to non-reciprocal motion when the oscillations are out of phase, leading to locomotion.
}
\label{collinear}
\end{figure}

A minimum of three particles is therefore needed to generate a net motion.
The simplest magnetocapillary swimmer is produced in the collinear configuration of 3 particles~\cite{grosjean2016}.
If we neglect the interaction between the two outermost spheres, we can consider the collinear assembly as a combination of two particle pairs, named $a$ and $b$ (see Fig.~\ref{collinear}).
Because the particles are in line, it is possible to only excite the radial mode in $d$, keeping the motion one-dimensional.
If the particle pairs are identical, they oscillate in phase and the motion is reciprocal.
However, a phase difference can appear between the oscillations $a$ and $b$ if each pair possesses a different stiffness, namely $k_{d,\:a} \neq k_{d,\:b}$.
This can be achieved experimentally by changing the diameter of one of the outer spheres~\cite{grosjean2016}.
When a phase difference exists between the oscillations of each pair, the non-reciprocal sequence observed is similar to the Najafi-Golestanian model for a minimal microswimmer~\cite{najafi2004,golestanian2008,pickl2012,pande2015}.
The speed of such a swimmer is a function of the amplitudes $A_a$ and $A_b$ of each oscillation as well as their phase difference $\Delta\phi$, such that
\begin{equation}
V = K A_a A_b \omega \sin(\Delta\phi)
\label{Vgol}
\end{equation}
where $K$ is given by the geometry of the system~\cite{golestanian2008}.
An analytical expression for $V$ can be found that is a function of viscous damping and the stiffness of the oscillators.
This simple linear model accounts for the speed profile observed in the experiments, as well as the observed speeds of the order of 0.02~$D/T$ where $T=2\pi/\omega$ is the period of the oscillating field~\cite{grosjean2016}.

The case of the triangular swimmer is more complex.
Indeed, all three particle pairs must be considered.
Furthermore, because of the geometry, both $r$ and $\theta$ modes will be excited regardless of the orientation of $B_h$.
The lack of simplifying symmetry and the non-linearity of the interactions at play make it difficult to reach any analytical formulation.
However, one can understand the origin of non-reciprocal motion when $\omega \rightarrow 0$ by studying the effect of a quasistatic variation of $B_h$.
While a hysteresis cycle was clearly identified in the transition between the triangular and collinear states, as shown in Fig.~\ref{configs}, this cycle is not the one used to produce swimmers in the experiment.
Indeed, a spontaneous transition from the triangular to the collinear state would only happen near the contact event, where any perturbation leads to the collapse of the swimmer.
A more practical way of producing swimmers takes advantage of the coexistence of two isosceles states.

\begin{figure}
\includegraphics[width=\linewidth]{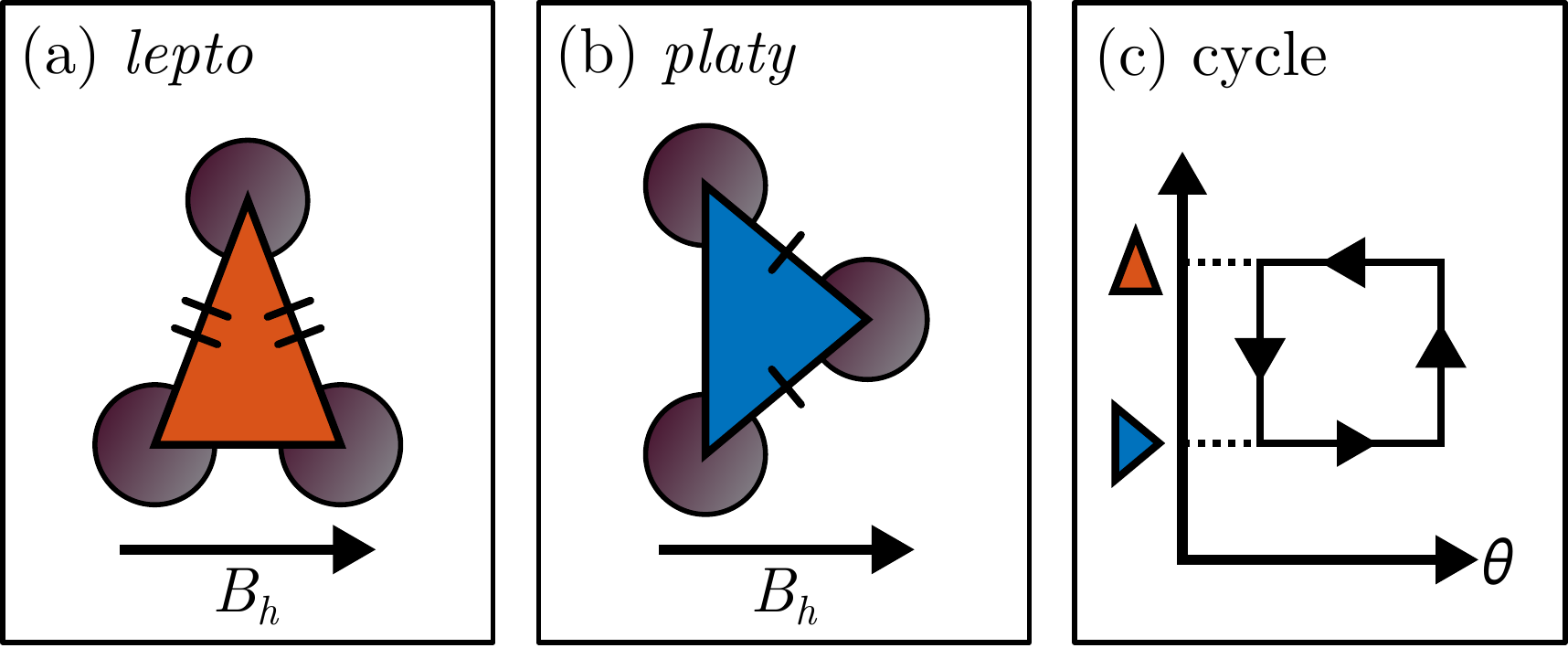}
\caption{
Two isosceles configurations can be observed in 3-particle assemblies.
The \emph{lepto} (a) has two internal angles above $\pi/3$ and the \emph{platy} (b) has only one angle above $\pi/3$.
The switch from one state to the other is accompanied by a rotation of the triangle, leading to non-reciprocal motion.
This is represented by a cycle (c) in the configuration space, where $\theta$ is the orientation between the triangle and the field $\vec{B}_h$.
}
\label{isosceles}
\end{figure}

Indeed, it has been shown~\cite{grosjean2015} that several isosceles configurations can coexist, depending on the orientation of the triangle.
One is a pointy isosceles, called \emph{lepto}, with two angles above $\pi/3$ and one below (see Fig.~\ref{isosceles}a).
In this case, two of the three particles are aligned with $B_h$.
A second configuration is a flat isosceles, called \emph{platy}, with two angles below $\pi/3$ and one above (see Fig.~\ref{isosceles}b).
In that case, two of the three particles are perpendicular to $B_h$.
Because of their similar shape, those states are not distinguished on Fig.~\ref{configs}a.
However, it was shown previously in a similar Monte-Carlo study~\cite{grosjean2015} that the \emph{platy} state coexists with the \emph{lepto} state up to $B_h/B_z \approx 0.35$, above which the \emph{lepto} is the only triangular configuration observed.
One can see from Fig.~\ref{isosceles}a,b that the transition between those states must be accompanied by a rotation of the structure.
However, transition and rotation do not happen simultaneously, producing non-reciprocal motion, as illustrated in Fig.~\ref{isosceles}c.
Note that, in contrast with the collinear swimmer, the triangular one can break time-reversal symmetry with three identical particles.

\begin{figure}
\includegraphics[width=\linewidth]{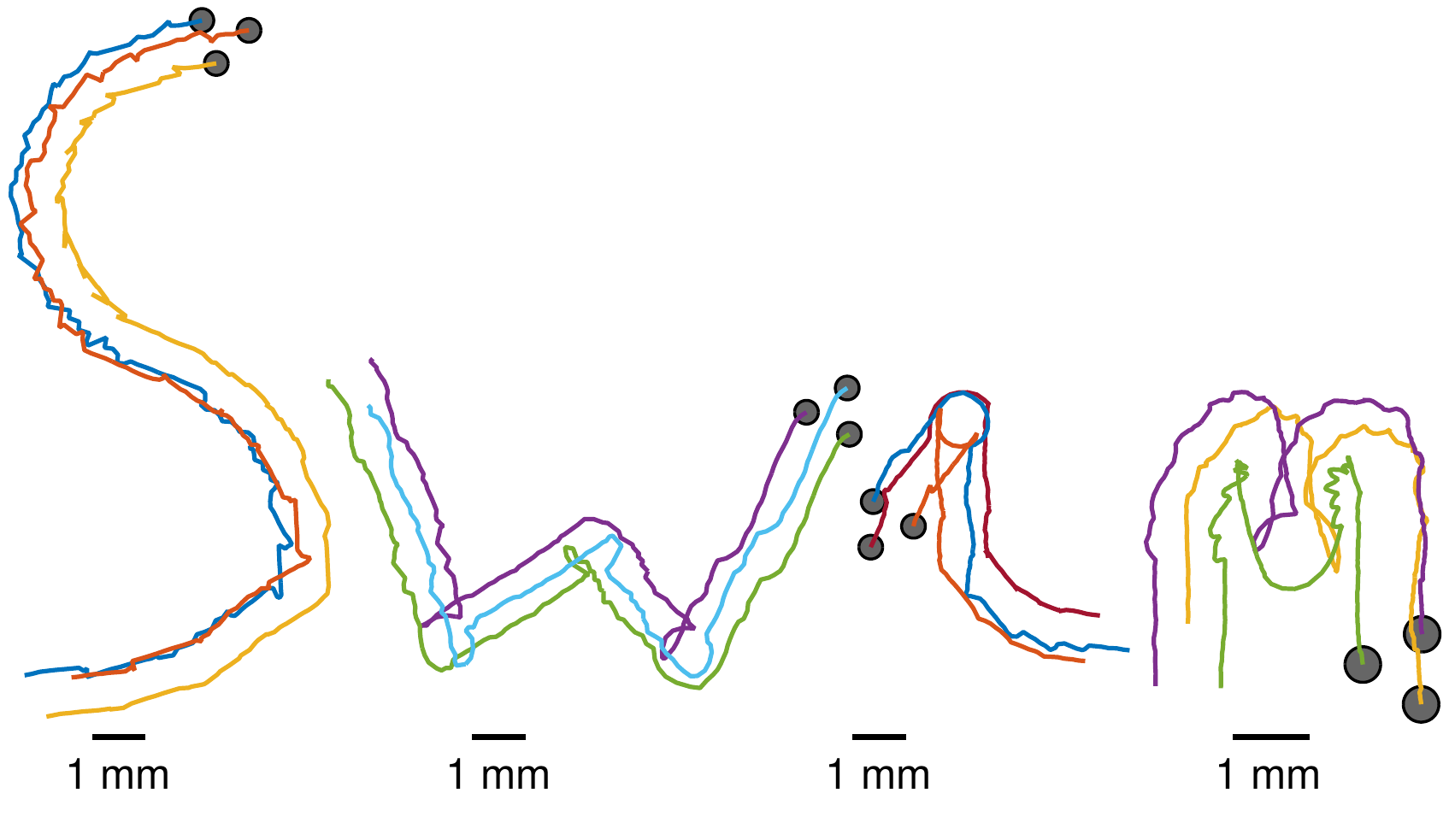}
\caption{
Remote-control of a three-sphere swimmer by adjusting the orientation of $B_h$.
Four trajectories, representing the letters ``Swim", are shown to illustrate the level of control achieved.
The letter ``S" shows a smooth, curved trajectory.
The ``w" illustrates a succession of straight lines and sharp turns.
The cursive ``i" is a combination of a smooth turn and a sharp reorientation.
Finally, the ``m" shows all of the aforementioned movements in a small region of space.}
\label{SWIM}
\end{figure}

One can wonder what is the direction followed by each swimmer.
In the case of the collinear swimmer, the swimmer moves in alignment with the field, in the direction determined by the sign of the phase difference $\Delta\phi$.
In general, however, swimming direction is at a non-trivial angle $\delta$ with the direction of $\vec{B}_h$.
If $\vec{B}_h = B_x \sin(\omega t) \vec{e}_x$, and granted that there is a unique $\delta$ between 0 and $\pi/2$, by symmetry, we can expect at least four swimming directions depending on the initial orientation of the assembly, namely $\delta$, $-\delta$, $\pi+\delta$ and $\pi-\delta$.
This is usually the case with the triangular swimmer.
Nonetheless, the trajectory of a triangular swimmer can be remote-controlled rather precisely, as shown on Fig.~\ref{SWIM}.
Indeed, once a swimming direction $\delta$ is observed, changing the orientation of $\vec{B}_h$ by an angle $\epsilon$ changes the swimming direction to $\delta+\epsilon$.
In order to facilitate the control, an offset $B_{h,\:0}$ is added to $\vec{B}_h$ so that it does not oscillate around zero any more.
A small offset of 1/10\textsuperscript{th} of the oscillation amplitude is enough to create a preferential orientation for the assembly on average, which helps keeping the swimmer in a well-defined swimming mode~\cite{grosjean2015}.
Compared to the collinear swimmer, higher speeds usually around $0.3~D/T$ are reached, with the fastest ones typically reaching $0.6~D/T$.
This relatively high speed explains why the triangular swimmer is used in remote-control experiments, despite the increased complexity.
For comparison, the early example of artificial microswimmer in~\cite{dreyfus2005}, which uses a magnetic filament for propulsion, achieved a comparable top speed of around $0.1~L/T$, where $L$ is the length of the filament.
Using the same definition for $L$, the speed of biological flagellates is typically quite low, of the order of $0.01~L/T$~\cite{derosier1998}.
This is explained by the very fast rotation of the flagellum and the prevalence of absolute speed over energy efficiency in this case~\cite{purcell1977}.

It is important to note that, while the quasistatic approach offers an intuitive explanation of the origin of reciprocal motion in the triangular case, it is not a definitive approach to characterize the swimmers.
Indeed, dynamical effects play an important role in the deformation of the assemblies.
For instance, no hysteresis loop is observed in a quasistatic deformation of a 4-particle swimmer, as was shown in Fig.~\ref{configs}.
However, although initially thought to be poor swimmers~\cite{lumay2013}, the 4-particle swimmers can reach similar speeds to the triangular swimmer, about $0.3~D/T$.
This shows the limit of the quasistatic approach and the necessity of studying the interaction dynamics in more detail.

\section{Applications}

Magnetocapillary swimmers have some unique characteristics that set them apart from other artificial microswimmers, which could be beneficial from the points of view of both fundamental and applied research.
Notably, they are bound to an interface, self-assembled from simple components, chemically inert and fuelless, while being relatively fast and controllable.
Naturally, many challenges remain to be addressed before any actual technological application can be considered for microswimmers in general, and magnetocapillary self-assemblies in particular.
Nonetheless, performing some simple tasks with the swimmer can serve as a proof of concept for future applications as well as a basis for further studies.
Several applications are commonly cited for microswimmers, notably concerning the manipulation and transport of micro-objects~\cite{dreyfus2005,hernandez2005,raz2008,lauga2009,zhang2010,baraban2012,tottori2012,sengupta2014,cheang2014,gao2014,ding2016} and the mixing and pumping of fluids~\cite{lauga2009,wu2000,kim2004,kim2007,pushkin2013,jalali2015}.
Other applications could make use of the magnetic properties of the spheres.
For example, it is possible to produce targeted heating in a fluid using magnetic particles under high frequency oscillating fields, which has long been studied for cancer treatment~\cite{giustini2010}.
The heating of a floating particle, in this case by a laser, has also been used to manipulate objects on a water surface by locally changing surface tension~\cite{mallea2017}.
This technique could be applied to magnetocapillary assemblies to provide greater control.
In this section, two experiments are performed and discussed in order to illustrate how magnetocapillary swimmers could perform two simple tasks : the transport of an object and the local mixing of fluids.

One of the main tasks proposed for microswimmers is the transport and delivery of a cargo~\cite{hernandez2005,raz2008,lauga2009,zhang2010,baraban2012,tottori2012,sengupta2014,cheang2014,gao2014,ding2016}.
Indeed, the controlled transport of a micro-object could potentially be used in microfabrication, manipulation of biological components or targeted drug delivery~\cite{zhang2010,cheang2014,gao2014,sitti2015,ding2016}.
The transport process can be divided into three steps: the capture, the towing and the release of the cargo.

\begin{figure}
\includegraphics[width=\linewidth]{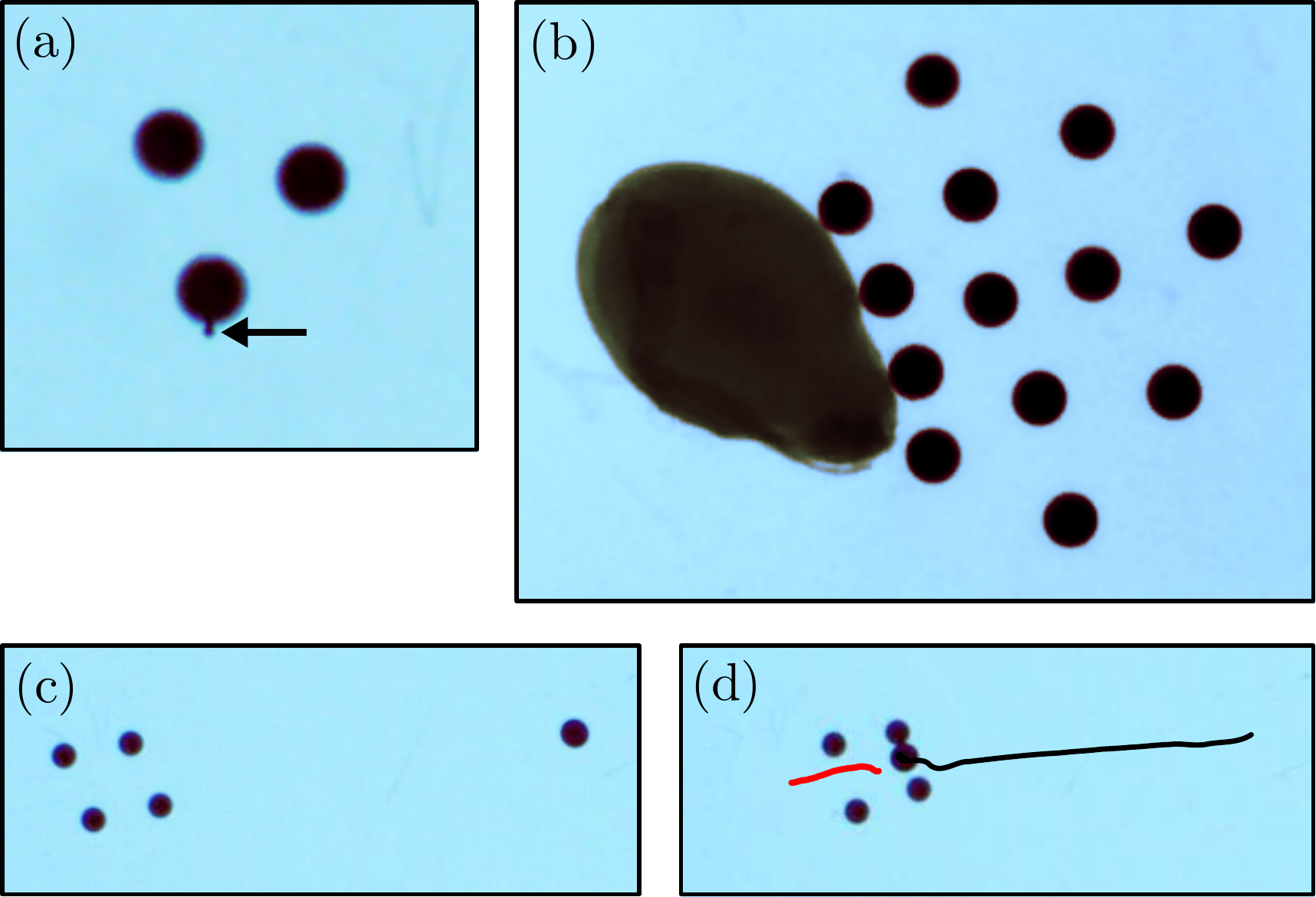}
\caption{
Capture of floating particles by capillary attraction.
Objects of various sizes can be captured, such as a polyethylene sphere of 80~$\mu$m (a) or a sesame seed of 3.7~mm (b).
(c) A 4-particle swimmer is brought in the proximity of a polyethylene sphere of about $600~\mathrm{\mu m}$.
(d) After about 15~s, the sphere is captured by the swimmer through capillary attraction.
Lines show the motions of their center of mass.
}
\label{capture}
\end{figure}

The capture of a floating object by a magnetocapillary swimmer can be relatively straightforward by using capillary forces, granted that the swimmer and the object possess like capillary charges.
In other words, in the proximity of the swimmer, the metallic spheres naturally attract objects that produce a concave meniscus, which is the case for most heavier than water objects.
The capture of several floating objects by swimmers made of 500~$\mu$m particles is shown on Fig.~\ref{capture}.
This capture process has been successfully tested on various particles sizes, including a polyethylene sphere of about 80~$\mu$m (see Fig.~\ref{capture}a) and a sesame seed about 3.7~mm long (see Fig.~\ref{capture}b).
Finally, the capture of a polyethylene sphere of 600~$\mu$m is shown in Fig.~\ref{capture}c,d.
Note that a particle having a capillary charge of the opposite sign as the swimmer would experience a repulsion and be pushed away from the swimmer in a way that would be difficult to precisely control.
Nonetheless, capture could still be achieved.
For example, the swimmer could capture an intermediary, amphiphilic particle, possessing both negative and positive capillary poles~\cite{vella2005,poty2014}.
Capillary gripping has been proven to work in other systems using bubbles~\cite{giltinan2014} or droplets~\cite{lambert2007} as intermediary, although this does not rely on the Cheerios effect.
Furthermore, by introducing defects in the contact line of the spheres, one could produce a directed capture at a specific site on the object~\cite{wong2016}.
Finally, it might be possible to change the sign of the capillary charge of the swimmer by using another body force opposing gravity, for example with a vertical magnetic gradient~\cite{tsai2013,vella2015}.

The main drawback of the capillary interaction as a capture method is its lack of selectivity.
Any particle possessing a capillary charge of the same sign as the swimmer will be attracted when it comes in proximity with the swimmer.
This does not present a problem, though, if the intended goal is to collect as many floating particle as possible, for example in pollution removal applications~\cite{vilela2016}.
A possible strategy to obtain a selective capture by capillary interaction would be to code a specific succession of capillary poles on the particles, by forming an undulated contact line or using non-spherical particles~\cite{loudet2005,loudet2006,danov2010,davies2014}, matching the code of the target particle.
However, most applications requiring the selective capture of a target or the capture of objects possessing capillary charges close to zero would necessitate other methods.
It has been shown that a cargo can be kept close to the swimmer during transport by hydrodynamic interaction alone~\cite{diller2014}.
Note that because magnetocapillary swimmers are chemically inert, and their locomotion is not based on a chemical interaction with their environment, coating the particles to obtain desired properties~\cite{gera2010} is an option.
For example, targeted delivery of genes using magnetic swimmers coated with a reagent has already shown promising results in the case of artificial bacterial flagella~\cite{qiu2015}.

The second step of the transport process is the cargo towing itself.
Depending on the size of the load relative to the swimmer, this can vary from a trivial matter to a very complex problem.
If $L$ denotes the typical length of the cargo, an object such that $L\ll D$, as in Fig.~\ref{capture}a, will have little influence on the deformation of the swimmer and the resulting fluid flow.
On the other hand, the case of a cargo much larger than the particles, \emph{i.e.} $L\gg D$, is significantly more complex.
The motion of several particles can be impeded by the cargo, which means that a larger number of particles is needed to produce non-reciprocal deformations.
For example, in Fig.~\ref{capture}b, 13 particles have been used to capture a 3.7~mm long sesame seed.
Finding efficient swimming regimes in assemblies of a large number of particles remains an open question.
Indeed, each subgroup of particles in the assembly would need to work in synergy, generating a net flow in the same global direction, in order to overcome the increased drag.
This is usually not the case in the experiment.
With the sesame seed of Fig.~\ref{capture}b, only low speeds of the order of $10^{-2}~D/T$ were observed.
The case of large cargoes should be further investigated, in particular to determine what the optimal number of particles is, depending on $L$ and $D$.

\begin{figure}[t]
\includegraphics[width=\linewidth]{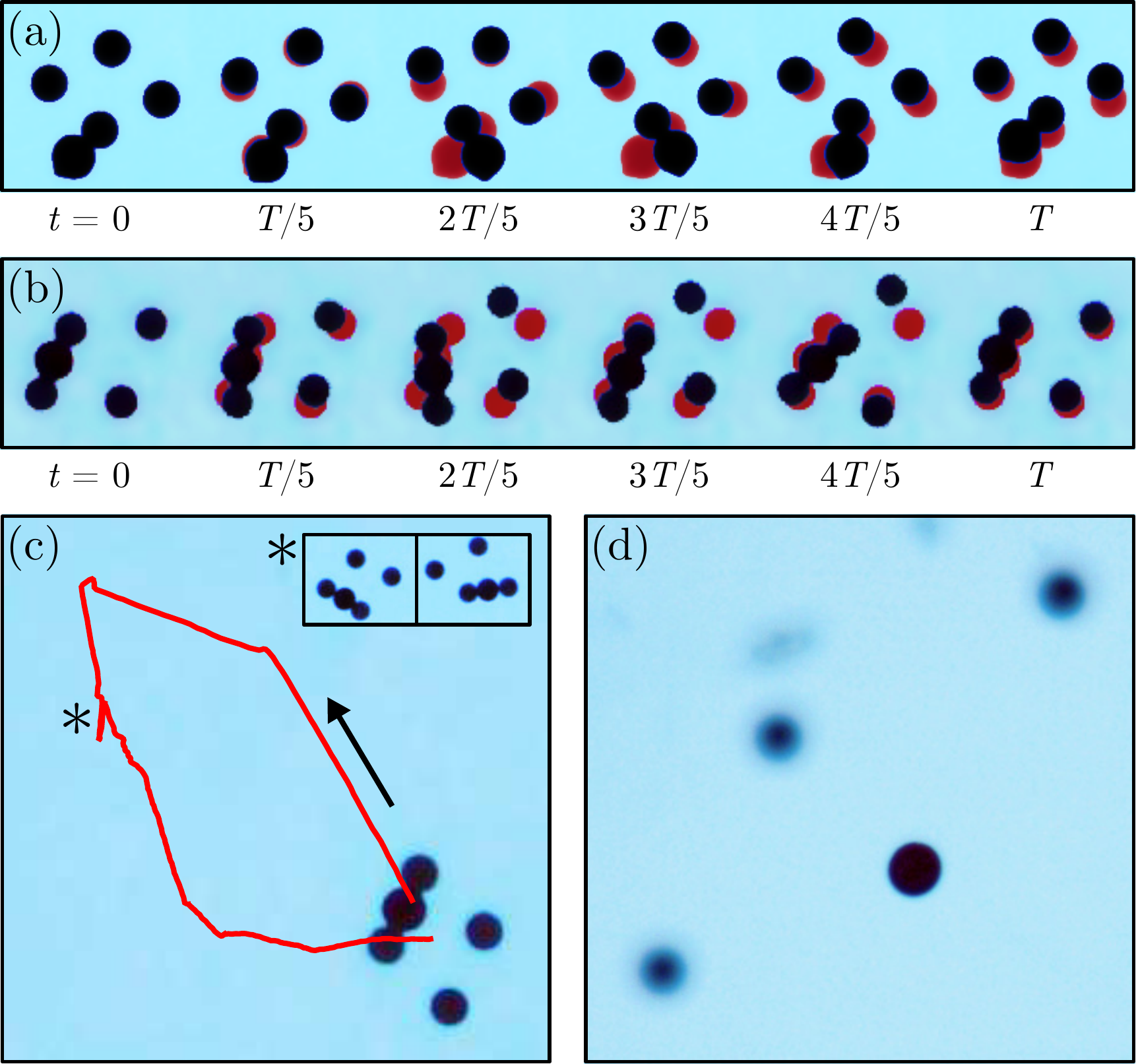}
\caption{
Towing and release of a floating particle.
(a) Under a sinusoidal field $B_h (t)$, the assembly and cargo swim.
Non-reciprocal deformations are shown over one period of the oscillating field $T=2~\mathrm{s}$.
(b) Same as in (a), following the capture shown in Fig.~\ref{capture}.
In this case, the cargo is stuck between two particles.
(c) Nonetheless, swimming direction is controllable.
Here, the cargo is moved then brought back to its initial position.
A notch in the trajectory (*) happened when the cargo jumped from one particle pair to another, as shown on the insert.
(d) The non-magnetic cargo is released by sinking the metallic spheres using a non-uniform magnetic field, \emph{i.e.} by approaching a neodymium magnet.
}
\label{transport}
\end{figure}

When $L\approx D$ (as in Fig.~\ref{capture}c,d) the cargo can more or less influence the motion of the particles depending on the situation.
Figure~\ref{transport}a shows the towing by a 4-particle swimmer of a 600~$\mu$m polyethylene sphere over one period of the oscillating field at 0.5~Hz.
The deformation of the swimmer is superimposed to the initial configuration at $t=0$, in order to highlight the displacement of each particle after each time step of $T/5$.
The load is in contact with a single magnetic particle.
Its motion evidences a rotation of the particle in question, showing that its movement is relatively unhampered by the presence of the cargo, in spite of the added mass and viscous drag.
It is not impossible that the presence of the cargo may actually improve propulsion, as evidenced by the relatively large swimming speed of about 0.5~$D/T$.
This might be an effect of the large amplitude of motion of the cargo.
Furthermore, it is has been shown numerically that an arbitrary swimmer loaded with a cargo can move faster than an unloaded one in some conditions~\cite{raz2008}.
Namely, an optimal size ratio between a swimmer and a cargo can exist where swimming efficiency is significantly improved, granted that the cargo be sufficiently close to the swimmer.

Figure~\ref{transport}b shows the towing of a similar polyethylene sphere, following the capture which was shown in Fig.~\ref{capture}c,d.
In this case, the sphere is in contact with two particles of the assembly.
This has the effect of creating a solid link between those particles, restricting their motion.
However, non-reciprocal motion is still observed, leading to a swimming speed of about 0.25~$D/T$.
In both situations, swimming trajectory is controllable.
Figure~\ref{transport}c shows the transport of the cargo on a small loop.
Every change in direction corresponds to a change in the direction of the oscillating field $\vec{B}_h$, with the exception of a small notch in the trajectory (*).
This notch corresponds to the moment the cargo detached from a particle and came into contact with another one.
Despite this perturbation, the swimmer was successfully brought back to its starting position.

The last step of a transport experiment is the release of the load.
If the cargo possesses a sufficiently large magnetic susceptibility, a strong vertical magnetic field will move the particles and cargo away from each other, freeing the load.
On the other hand, if the magnetic susceptibility of the cargo is small compared to which of the particles, as is the case with the polyethylene sphere, a strong vertical magnetic gradient can sink the metallic spheres while the cargo stays afloat.
This was done in Fig.~\ref{transport}d by approaching a neodymium magnet from below the bath.
The sphere in the foreground is the polyethylene cargo, while three of the four particles constituting the swimmer can be seen in the background, out of focus.

\begin{figure}[t]
\includegraphics[width=\linewidth]{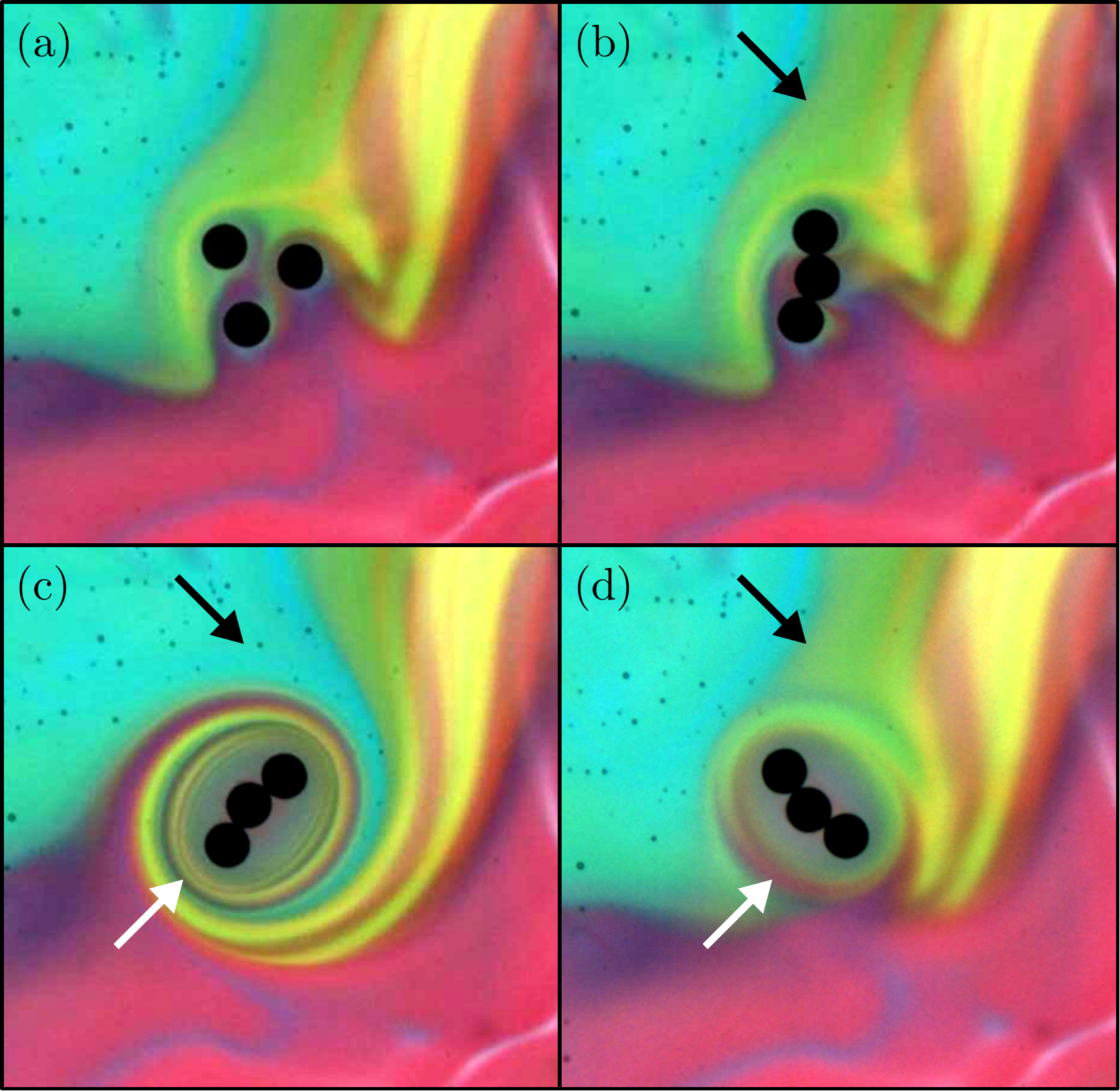}
\caption{
Illustration of a simple mixing device.
(a) A triangular swimmer is moved to the intersection of three fluids, each a glycerol-water mixture with a red, yellow or blue dye.
(b) A large $\vec{B}_h$ is used so that the particles collapse in a line.
(c) $\vec{B}_h$ is rotated for about 8 turns clockwise.
Fluid is entrained as the line follows the rotation of $\vec{B}_h$.
(d) $\vec{B}_h$ is rotated for about 8 turns counter-clockwise.
Far from the particles, the flow is reversible (black arrow).
Close to the particles, however, mixing has occurred due to molecular diffusion (white arrow).
}
\label{mixing}
\end{figure}

Another commonly cited application of microswimmers is mixing at the micro scale~\cite{lauga2009,wu2000,kim2004,kim2007,pushkin2013,jalali2015}.
For instance, swarms of microswimmer can enhance the diffusion of constituents in a fluid~\cite{wu2000} or the mixing of several fluids in microfluidic devices~\cite{kim2004,kim2007}.
In general, a microswimmer generates fluid motion over a long range, producing stirring.
This stirring effect is most pronounced if swimming direction is frequently changed in a random way, in a so-called run and tumble motion~\cite{lin2011,pushkin2013}.
In the case of a remote-controlled artificial microswimmer like the magnetocapillary swimmer, parameters such as the typical run length between reorientations, the distribution of run lengths and the smoothness of the turns can be varied.

Another possibility is to generate local stirring by rotational motion or motion on a closed loop~\cite{jalali2015}.
This could prove useful in confined environments, such as small droplets or microfluidic devices, where stirring by run and tumble motion loses efficiency~\cite{pushkin2014}.
Furthermore, assembling an array of rotating objects has been shown to be a viable strategy for mixing in microfluidic devices~\cite{lu2002}.
Indeed, like the Reynolds number, the relative importance of advection and diffusion in a fluid, called the P\'eclet number, is proportional to the typical length scale and velocity of the flow.
Therefore, at the micro scale, molecular diffusion becomes the preferred mixing mechanism over advection.
In this case, the volume of fluid mixed can be defined as
\begin{equation}
V=\mathcal{A}\sqrt{Dt}
\label{volmix}
\end{equation}
where $D$ is the mass diffusion coefficient and $\mathcal{A}$ the contact area between the fluids to mix~\cite{lu2002}.
Maximizing contact area $\mathcal{A}$ is therefore a possible mixing strategy.
With this aim in mind, an investigation of the rotational regimes of magnetocapillary swimmers and their possible use as a micromixing device could prove useful.
The simplest way to produce rotational motion in a magnetocapillary swimmer is to submit it to a strong magnetic field rotating in the horizontal plane, causing the particles to come into contact and rotate.
In Fig.~\ref{mixing}, a triangular swimmer has been brought to the intersection of three fluids, each being the same mixture of water and glycerol with the addition of a red, a yellow and a blue dye.
A strong horizontal field $\vec{B}_h$ causes the particles to come into contact and form a line.
This magnetic rod follows the orientation of $\vec{B}_h$ in the manner of a compass needle.
After about 8 clockwise turns of the magnetic field, fluid close to the swimmer has been entrained, forming a spiral.
Contact area $\mathcal{A}$ has increased, as thin layers of each colour were intertwined (see white arrow on panel (c)).
The inverse rotation of $B_h$ is then applied to rotate the rod about 8 turns counter-clockwise.
On a circle of roughly the size of the swimmer, mixing by molecular diffusion has occurred, as evidenced by the grey area around the particles (see white arrow on panel (d)).
Further away from the swimmer, the fluid is back to its initial position due to the reversibility of laminar flows (see black arrows).
This simple mixing device illustrates the potential of artificial microswimmers to produce mixing at specific locations.
Note that, similarly to the swimming regimes, a rotational motion of a magnetocapillary swimmer can also be obtained from the non-reciprocal motion of the particle.
This should be further investigated due to the potential of creating more complex flows, as each particle follows an individual orbit, as opposed to the solid rotation shown in Fig.~\ref{mixing}.

\section{Conclusion}

In summary, combining the capillary interaction between floating objects with an interaction between magnetic induced dipoles leads to the formation of ordered structures by self-assembly.
Depending on the number of particles in the assembly and the amplitude and orientation of the imposed magnetic fields, a variety of structures can be observed.
The adopted configurations can also depend on the history of the system, in other words, hysteresis is observed.
This rich energy landscape partly explains the appearance of non-time-reversible deformations under periodic variations of the applied fields.
Some of these deformation modes lead to a net motion at low Reynolds number.
Furthermore, the swimming trajectory can be controlled in the plane of the interface.

A variety of applications can be envisioned for controllable microswimmers, based on their ability to move fluid and interact with micro-objects.
Examples include drug delivery, manipulation and assembly of small components, mixing and pumping.
This has been illustrated here, first by the capture, controlled transport and release of a floating cargo.
The capture is made by capillary interaction while the release of the cargo is achieved by sinking the swimmer with a magnetic gradient.
As is, this method is limited to cargoes of like capillary charge as the swimmer.
Furthermore, cargoes significantly larger than the particles of the swimmer can impede their movement.
To overcome this, many-particle assemblies must be used, which should thus be studied in more detail.
Indeed, changing the number of particles in an assembly can drastically change the equilibrium configurations and the dynamics.
A second possible application, the mixing of fluids, has been illustrated.
The rotation of a swimmer entrains fluid, intertwining the components to mix.
Such a process could be used to generate local mixing at a precise location, or to mix larger quantities by combining several mixers.
The flow produced by the rotation of various assemblies should be further studied with this aim in mind.

\section*{Acknowledgements}
This work was financially supported by the University of Li\`ege (Grant No. FSRC 11/36). GG thanks FRIA for financial support.

\section*{References}

\bibliography{biblio}

\end{document}